\documentclass[12pt]{article}
\usepackage{latexsym}
\usepackage[english]{babel}
\usepackage{fancyhdr}
\usepackage[mathscr]{eucal}
\usepackage[dvips]{graphicx}
\usepackage{graphics}
\usepackage{color}
\usepackage{epsfig}
\usepackage{amsmath}
\usepackage{amsthm}
\usepackage{amsfonts}
\usepackage{amssymb}
\usepackage{amscd}
\usepackage{bbm}
\usepackage{latexsym}
\usepackage{marvosym} 
\usepackage{pifont}
\usepackage{subfigure}
\usepackage{psfrag}
\usepackage{caption}

\begin{document}

Zero range interactions in d=3 and d=2 revisited 

\vskip 6 pt  \noindent 

\noindent
G.F:Dell'Antonio

\vskip 4 pt 

\noindent
Mathematics dept. University Sapienza (Roma)

and 

\noindent
Mathematics Area   Sissa (Trieste)

\section{Introduction}

This paper has a two-fold purpose: 

1) to clarify the difference between contact and weak-contact interactions (called point interactions in [A] in the case $N=2$) in three dimensions and their role in providing spectral properties and boundary conditions.

2) to analyze contact interactions in two dimensions. 

\bigskip
Both contact and weak-contact are "zero range interactions" or equivalently self-adjoint extension of $\hat H_0$,  the free hamiltonian for a system of $N$ particles, restricted to functions that vanish in some neighborhood of the \emph{contact manyfold} $ \Gamma = \cup_{i,j}\Gamma_{i,j}, \;\;\gamma_{i,j} \equiv \{ x_i - x_j = 0\} \;\; i \not= j =1 \ldots N $. 

They differ "for the quality of contact" and for the fact that weak-contact interactions require the presence of a zero energy resonance. 

Both correspond to idealized setting, as both can be obtained as limit, in the strong resolvent sense, of interactions due to central potentials of very short range; in the case of contact they are scaled as $ V^\epsilon (|y|)= \frac {1}{ \epsilon^3} V( \frac {(|y|)}{ \epsilon} )$ while in the weak contact case the scaling is 
$ V^\epsilon (|y|)= \frac {1}{ \epsilon^2} V( \frac {|y|}{ \epsilon} )$.

In the contact case for $N=2$ this limit must be taken in the Heisenberg representation (\emph{as generator of  groups of  automorphisms  of the observables} );  in the hamiltonian formulation the hamiltonian contains  a c-number term that diverges as $ \epsilon \to 0$

We show that when a combination of the two is used  there is no interference in the results.

This suggests, in the hamiltonian formulation, to use weak-contact interactions only in case the "contact" is obtained artificially through a zero energy (Fesbach) resonance and to use both when one wants and distinguish between the two, as is the case for Cooper pairs (since the range of one of them is small but "large as compared to the other's"). 

For  the spectra of "natural" N-body systems  with very short range interactions one should use  contact interactions as a model.

Indeed the structure of N body systems, in particular the presence of \emph{Efimov trimers} [E] and \emph{Efimov quadrimers},  is correctly predicted by the model "contact interactions"  (the name Efimov refers to the presence of an infinite number of bound states with energy levels that decrease geometrically).

\bigskip
\emph{Remark}

Notice that the "potential" in the weak-contact case has the same behavior as the laplacian under the dilation group. 

A zero mass particle  can be mapped under the action of this group into a zero energy resonance. 

Therefore weak contact can be seen also as contact with a zero mass particle. 

Since the hamiltonian  diverges when the mass of a particle goes to zero, the hamiltonian in the weak contact case is not defined (but the interaction term  and the equations for the interacting system are well defined).

\bigskip

............................

\bigskip

We consider briefly also the case of two space dimensions.

Also in two dimension the two types of "zero range interactions" can be defined 

The scaling for the contact interaction  is now $ V^\epsilon (|y|)= \frac {1}{ \epsilon^2} V( \frac {|y|}{ \epsilon})$  and the scaling for the weak  contact is $ V^\epsilon (|y|)= \frac {1}{ \epsilon} V( \frac {|y|}{ \epsilon})$,

In two dimensions one can  consider  a system $S$ of three particle  of mass $m$ in which two particle have  a contact interaction and a simultaneously a weak contact interaction with  a 
the third  particle  (the potential  is the product of two weak-contact potentials). 

This system  is described by a hamiltonian with  no zero energy resonance.

If one take one takes $ m = \frac {1}{\epsilon} $ where $ \epsilon $ is the parameter used to define weak contact as limit of regular potentials,  the limit $\epsilon \to 0$  leads to an increase without bound of the number of bound states while their energy decreases to zero.

In the limit  $ \epsilon \to 0 $ the system may be regarded as a \emph{point with internal structure}; the hamiltonian is positive and its spectrum has an essential singularity at  zero.  

The Wave operator for the interaction with a fourth particle extends to a bounded map on $L ^p $ for all $ 1 < p < \infty $ [E,G,G] \bigskip

\emph{Remark} 

It is convenient to take into account the following facts that may serve as guide for the zero range interactions. 

In three dimensions the potential for  weak-contact has the same transformation properties as the laplacian under dilations and therefore the zero energy resonance may be seen as a mass zero particle.

In three dimensions the contact potential has a different scaling under dilations from the laplacian;  therefore contact interactions  can be used only in the three body problem.

Contrary to what happens in three dimensions, in two dimension the contact potential has the same transformation properties under dilations as the laplacian and  can be used  in a two-body system.

In two dimensions in a three body system one can introduce \emph{the product of two weak-contact potentials}. 

No zero-energy resonances are produced and the interaction has the singularity of a contact interaction. 
 
\bigskip

..................................

\section{Contact and weak-contact interactions in three dimensions }

Recall that contact interaction are limits os $ \epsilon \to 0$ of regular two-body interaction with central potentials $V^\epsilon$  that scale as 

\begin{equation} 
V^\epsilon (|y|)= \frac {1}{ \epsilon^3 } V( \frac {|y|}{\epsilon} )
\end{equation} 

Weak contact interactions are limit, in strong resolvent sense, of two-body interactions with central potentials that scale as 

\begin{equation} 
U^\epsilon (|y|)= \frac {1}{ \epsilon^2 } U( \frac {|y|}{\epsilon} )
\end{equation} 

provided the two body hamiltonian $ H_0 + U_\epsilon$ has a zero energy resonance (a solution of $ (H_0 + U^\epsilon ) \psi (x) = 0$ that has an asymptotic behavior $ \psi (x) \sim \frac {1}{ |x|}  $ when $ x \to \infty$ ($x$ is the relative coordinate).

\bigskip
\emph{Remark} 

Contact interactions correspond to a distributional potential on $\Gamma_{i,j} \equiv \{x_1 = x_j,\} \; i \not= j$. 

From   (2)  one sees that the scaling properties  of the weak  contact interaction are the same as the scaling properties of the primitive (with respect to the radial variable) of the potential of  contact case (delta distribution) . 

Since these are only formally "potentials" it will be convenient to use an (invertible) map to  a space of more singular functions where the potentials are represented by more regular functions and perform the analysis there.

In this space resolvent convergence for  the approximating hamiltonians  has an easier proof.   

 \bigskip
 
 ...............................
 
 \bigskip
 
 Weak-contact were introduced in [A] (under the name of point interaction) for the case $N=2$ to be able to define an interaction localized in a point. 

In order to have an hamiltonian equation (and not only a group of automorphisms of of observables) the two-body hamiltonian must have  a zero energy resonance.  

For more that two particles contact and weak contact interactions are well defined in $ L^2 (R^{3N}) $ and one can use any combination of the two. 

We shall prove that when they are both used in the same system, they do not interfere (the outcome is the same as if they were applied separately).

\bigskip
Contact interaction may have an Efimov [E] sequence of three and four body bound systems .  

When $ N \geq 3 $ quasi-contact interactions (the name is chosen because in this case the decrease of   the support of the approximating potentials is slower) can give bound states due to \emph{conspiracy of a  zero energy resonance with a zero range interaction} (in the regular case the presence of an Efimov sequence is due to  conspiracy between  two-body resonances [A,S]).

Contact and weak-contact interactions can be present \emph{simultaneously} .

We will prove  that contact and weak-contact interactions produce , when they are simultaneously present,  \emph{complementary} and \emph{independent}  effects.

As a consequence a \emph{zero range interaction} is completely defined by giving \emph{the ratio between the contact and the weak-contact components}. 

In $R^3$ the contact interaction provides the Ter-Martirosian Skorniakov [S,T] boundary conditions at $ \Gamma $ that are used in Nuclear Physics (in the Fadeev formalism) and are also responsible for   the presence of  an Efimov sequence of trimers and quadrimers. 

Weak-contact interactions add a constant term in the boundary behavior 

If the interaction are both of weak-contact type the  hamiltonian,  depending on the masses and the coupling constants, may have  low energy bound states. 

\bigskip
\emph{Remark} 

In order to define weak-contact interaction in $R^3 $ between two particles \emph{as hamiltonian system} for an isolated system one must require $N\geq 3$.

In presence of an external magnetic field one can define weak contact interactions  but the the pace of $H_0$ is taken by the magnetic Scrh\"odinger  operator.

In the two particle  case  weak contact interactions are generators of  groups of automorphisms of observables; the hamiltonian is defined up to an (infinite) constant.

Contact interaction can be used in dimension 3 in the study of the Gross-Pitayeskii limit for a gas of identical bosons.

In this limit one considers configurations in which two particles are at a very short distance (in the limit they are \emph{in contact})  and  on the scale considered  the remaining particles are so far apart that  their presence can be neglected. 

Since the system has more than two particles, contact interactions are well defined as hamiltonian system.

The dynamics of each member of the pair feels the influence of the other member through a potential $V^\epsilon$  that in the limit vanishes if the particles are not in contact (it is a distribution on the contact manyfold).

It is natural to assume that  in the limit the interaction with a particle be  proportional to the probability that the particle be there.  

For identical particles this leads to a cubic focusing Schr\"odinger equation which is of hamiltonian type.

\bigskip

............. 

\section{Short review of the analysis in [D] of contact interactions}

We review briefly the analysis done in [D]. We assume $ N \geq 3 $.

Contact interactions are a class of self-adjoint extensions of the operator $ \hat H_0$ that represents the free hamiltonian of a system of $N$ particles in $R^3 $ of mass $ m_1, \ldots m_N$ restricted to functions that vanish in a neighborhood of the \emph{contact manyfold} 

\begin{equation}
\Gamma \equiv \cup_{i \not= k} \Gamma_{i,k} \qquad \Gamma_{i,k} \equiv  \{ x_i - x_k \} = 0 ,\;\; i \not k = 1, \ldots N 
\end{equation}

Both operators (kinetic energy and potential energy)  define  quadratic forms on continuous function in $L^{3N}$. 

The quadratic form of $H_0 $ is positive and the quadratic form of the potential is negative 
(we have assumed that the potential is negative).

Depending on the masses and the "intensity" of the interaction the sum of the quadratic forms is a positive or is indefinite (the form  is always positive for functions which have support away from $ \Gamma$). 

For  the approximating hamiltonians with potentials $ V^\epsilon ,\;  \epsilon > 0$ it is known that the quadratic forms have an extension to $ { \cal H}^1 $ as Dirichlet forms, since the potentials $V^\epsilon $ are of Rollnik class for all $ \epsilon > 0$. 

If the sum  of the forms is positive, it can be closed and the closure defines a closed positive quadratic form and therefore a self-adjoint operator, extension of $\hat H_0 $.

This operator is the limit, in strong resolvent sense, of the hamiltonian with the potentials $V^\epsilon$.

If the total form is not of definite sign  and not closed,  it has in general many closures  and these correspond to different self-adjoint extensions. 

To find  extensions one can follow  the analysis made by Birman, Visik and Krein [B][K] for positive operators and choose for the extension the domain of $ (\hat H_0)^* $.

We follow an alternative way and  perform a compact invertible immersion of  $L^2 (R^{3N})$ into a larger space  Hilbert space and there make use of the theory of quadratic forms as discusses e.g. in [K,S]. 

The embedding is done using the (compact, invertible) operator $ ( H_0 + \lambda)^{- \frac {1}{4} } $ 

The quadratic forms are now defined as operators in $ { \cal H}^{\frac {1}{2} }$.

In [D]  this map is called  \emph{Krein map}  and indicated with $ { \cal K}$. The target space  is  called $ {\cal M}$ \emph{Minlos space}. 

The map could also be called "Friedrichs map"  because of the similarity with of the analysis of  the laplacian on $[0, + \infty)$

\bigskip

In $ { \cal M}$ the operator $H_0$ is represented by is square root (the forms are continuous in $ \lambda $ in this space and one can set $ \lambda =0$) .

The map takes the boundary potential at $ \Gamma_{i,j} $ into  a $ L^1 $ function which is limit, in this norm, of \emph{ the images under the Krein map} of the approximating potential $V^\epsilon$. 

The fact that in $L^2 (R^{3N})$  the total hamiltilonian is not self-adjoint and has a family of self-adjoint extensions is seen in the space  $ {\cal M}$ as an instance of the Weyl limit circle ambiguity.

If  the masses are very different and/or  if the potential is strong enough  these operators are  unbounded below and  show  the Thomas effect  (geometric divergence  of the eigenvalues to $- \infty$).

For weak- contact interactions, the image  of the potential  on $ \Gamma_{i,j}$ is proportional to  $log (|x_i - x_j | )$ and may produce a finite number of bound states.  

One can return now  to "physical space.  

In the contact case one has still a family of self-adjoint extensions as limit set but now, due to the change in metric, each family shows to Efimov effect (geometric converge of the eigenvalues to zero).

One can decompose the forms using invariance under rotation (the central potential and $H_0$ are rotation invariant). In each "sector" the form of each of the extension is strictly convex and so is their union. 

The convex set of limit forms is in the closure of the forms associated to the potentials $ V^\epsilon $;  compactness is assured since  the limit forms are bounded  in $ {\cal H}^1 $. 

 Gamma-convergence [Dal] provides a unique minimum and therefore a \emph{unique} limit under sequential convergence of the (Dirichlet) forms associated to the approximating  potentials $V^\epsilon$

Gamma convergence of quadratic form implies strong  resolvent of the associated operators [Dal]

In the weak-contact case, in "physical space" the image is still a self-adjoint operator, possibly with a finite number of bound states, and  there is no need to use variational techniques. 

Strong convergence of the approximating potentials implies strong resolvent convergence of the limit operators. 

\bigskip

\emph{Remark} 

For contact interactions this is \emph{an interpretation} of the steps taken by R.Minlos in  [M1, \: M2].

A \emph{different  interpretation} corresponds to weak-contact interactions (see e.g. [C]).

In this second interpretation  the hamiltonian is much less singular. 

The domain of the Hamiltonian contains the symmetrized product of the two zero energy resonances; the hamiltonian takes value zero on this vector. 

As in the contact case we  map back  from  Minlos space the quadratic inverting the Krein map. 

Since the potential is much less singular in the weak-contact case the resulting system is much more regular. 

Inverting the Krein map one has \emph{in physical space} a self-adjoint operator that for a proper choice   of masses and strength of the interaction may have  bound states.

\emph{But in doing so one does not have the Ter Martirosian boundary conditions at $ \Gamma$  (the functions take a finite value at the boundary) and one loses the Efimov effect.}

We will prove that these two schemes (both of which we have attributed  to Minlos)  \emph{do not interfere} and one may have both contact and weak-contact at the same time.

This  provides a hamiltonian with  boundary condition $ \frac {C_{i,j} } { |x_i - x_j|} + D_{i,j} $ at $ \Gamma_{i,j}  $ for the functions in its domain and may have an Efimov effect.. 

The \emph{ratio} of the coefficients $ C_{i,j} $ and $ D_{i,j} $ \emph{characterizes completely} the self-adjoint extension.
 
If both contact and weak-contact are introduced for the  study ofscattering problems,  the Wave operator is the product of the two Wave operators.

\bigskip
....................................

\section{Contact vs weak-contact}

The fact that in the proof of uniqueness of contact interaction we  use variational arguments and a minimum principle leaves room for the existence of other critical points and therefore other self-adjoint extensions. 

Since the parameters of the sequences were the size and support of the approximating potentials there may be other choice of these sequences that give different limits. 

In particular there may be choices for which one has sequential convergence \emph{from below} to the \emph{least negative limit-quadratic form}. 

We saw that different zero range interactions are obtained by different choices of the rate of contraction  for $ V_1$ and $V_2 $

\begin{equation}
V^\epsilon_1 (|x|) = \frac {a}{\epsilon^3}  V( \frac {|x|}{ \epsilon})
\qquad 
V_2^\epsilon =  \frac {b}{\epsilon^2 } V(\frac { |x|} {\epsilon} )
\end {equation}

Or by a choice of a rate of contraction for different potetials.

One can even for each potential take the weighted sum of the two rates.

For a system of particles with two-body  zero range interactions  the maximal negative quadratic form is obtained by choosing \emph{contact interactions} for any three-body interacting system , the minimal   one by choosing  \emph{weak-contact interactions}.

In  the first case, if the negative part of the spectrum is non empty, there are Efimov sequences of trimers and possibly  quadrimers.

The first case is in line with our variational analysis (in which the lim-inf of convex functionals is considered.

The second case (weak-contact interactions) corresponds to the choice of  the lim. sup. 

This could be done by a variational analysis, considering resolvents rather than quadratic forms.

In a many particle system one can made either choice for any pair of zero range interactions.

The limit is then a mixture of contact and  weak-contact interactions. 

This leads e,g, in a four particle system to the \emph{simultaneous presence} of contact and weak-contact interactions  (between different pairs), as is the case for Cooper pairs.

We stress once more that the presence of both interactions gives results that do not interfere. 
 \bigskip
 
 ............................

\section{A unified approach }

\bigskip
\emph{Theorem 1}

In three dimensions for $ N \geq 3 $   contact interactions and  weak-contact interactions  contribute \emph{separately} and \emph{independently}  to the spectral properties and to the boundary conditions at the contact manyfold.

Contact interaction contribute to the Efimov part of the spectrum and to the T-M boundary condition
$ \frac {c_{i,j} }{|x_j - x_i|} $ at the boundary $ \Gamma \equiv \cup_{i,j} \Gamma_{i,j} $. 

Weak-contact interactions contribute o the constant terms  on the boundary and   
may contribute to the  negative part spectrum.

\hfill $ \diamondsuit $

 \bigskip
 ..................................................

\bigskip
\emph{Remark} 

This theorem states  that all results of the  weak-contact case (in particular for point interactions) remain valid when contact interactions are added. 

Weak-contact refer to the case of potentials of very short range accompanied by a zero energy resonance.

Contact interaction refer to potentials  of "still shorter" range so they can be approximated by a delta distrbution  at the boundary.

When $N \geq 3 $  contact and weak-contact interactions are (different) extensions of the free hamiltonian $ \hat H_0 $ defined on functions that vanish in a neighborhood of the boundary $ \Gamma$.

Both contact and weak-contact interactions are limits when $ \epsilon \to 0$, in the strong resolvent sense, of smooth central potentials with radius of order $ \epsilon $ (and  different scalings).

\bigskip

..........

\bigskip

\emph{Theorem 2}

In three dimensions for $ N \geq 3 $    weak-contact interactions and interactions through regular potentials contribute  \emph{separately} and \emph{independently}  to the spectral properties. 

Interactions through  regular two-body potentials do not change the value of the wave function at $ \Gamma$ but modify  the spectral measure.

Weak-contact interactions contribute to the value of the wave function at $ \Gamma$ and may  originate   bound states. 
  
\hfill $ \diamondsuit  $ 
 
\bigskip

......................................

\bigskip

\emph{Remark} 

It is interesting to notice  that we always make  reference to \emph{potentials} as is usually done in Quantum Mechanics. 

If one consider \emph{forces}, i..e gradient of potentials, it is the weak-contact case that leads to forces as distributions  supported by the boundary $\Gamma$ while in the contact case one cannot define \emph{boundary forces}. 

Therefore the weak-contact interactions should be used  when considering a semi-classcal limit:
recall that in classical hamiltonian mechanics, the \emph {constraining forces} are concentrated at the boundary.

\bigskip
..........................

\bigskip
For an unified presentation  is convenient to use a symmetric presentation due to Kato and Konno-Kuroda [KK]  (who generalize previous work by Krein and Birman)  for hamiltonians that can be written in the form

 \begin{equation}
H= H_0 + H_{int} \qquad H_{int} = B^* A   
\end{equation}

 where $B, \;A$ are densely defined closed operators with $ D(A) \cap D(B) \subset D(H_0) $ and such that, for every $z$ in the resolvent set of $H_0$, the operator $A \frac {1}{ H_0 + z} B^* $ has a bounded extension, denoted by $Q(z).$
 
We give details in the case $N=3$.

Since we consider the case of attractive forces, and therefore negative potentials it is convenient to denote by $ - V_k( |y|) $ the two body potentials

The particle's  coordinates are $x_k \in R^3 , k=1,2,3$ 

We take the interaction potential to be 

\begin{equation} 
V^\epsilon (X) = \sum_{i \not= j} [ V_1 ^\epsilon (x_i -x_j) +V_2 ^\epsilon (x_i-x_j)  + V_3 ^\epsilon (x_i-x_j)]
\end{equation} 

where $V_k(|y)| $ are regular potentials,.

For each pair of indices $ i,j$ we define   $ V_1^\epsilon (|y|) ) = \frac {1}{\epsilon^3}  V_1(\frac{|y|}{\epsilon})$ and $ V_2 ^\epsilon (|y|) = \frac {1}{\epsilon^2}  V_2 (\frac{ |y|}{\epsilon})$.

We leave $ V_3 $ unscaled. 

The limit correspond respectively to contact , weak-contact and regular perturbations. 

We define $ B^\epsilon = A^\epsilon = \sqrt { V^\epsilon}$.

 For $ \epsilon > 0$ using Krein resolvent formula one can give explicitly the operator $B^\epsilon $ as \emph{convergent power series} of products of the free resolvent $ R_0 (z) , Re z > 0 $ and the square roots  of the potentials $ V_k  ^\epsilon \ \   k=1,2,3$.
 
 One has then for the resolvent $ R(z) \equiv \frac {1}{ H +z } $  the following form [K,K] 

\begin{equation}
R(z) - R_0 (z) = [R_0(z) B^\epsilon ] [1 - Q^\epsilon(z )]^{-1} [ B^\epsilon  R_0 (z)]   \qquad z > 0
\end{equation}  

with

\begin{equation}
R_0 (z) = \frac {1}{ H_0 +z} \qquad Q^\epsilon (z) = B^\epsilon \frac {1}{ H_0 + z }   B^\epsilon \qquad  B^\epsilon= \sqrt { V^\epsilon (x) +V^\epsilon (y) }
\end {equation}

\bigskip
\emph{Proof of Theorem 1 }

The $L^1$  norm of $ V_1^\epsilon$ and the  $L^2 $ norm of $V_2 ^\epsilon $ do not depend on  $ \epsilon $.

We approximate the zero range hamiltonian with the one parameter family of hamiltonians 

\begin{equation} 
H_\epsilon =  
 H_0 + + \sum_{m,n}V^\epsilon (|x_n -x_m|)  \;\; n \not=  m , x_m \in R^3 
 \end{equation}

The potential is a the sum of three terms 

\begin{equation} 
V^\epsilon (|y|) = \sum_{i=1}^3  V_i ^\epsilon, \quad   V_1^\epsilon (|y|) =  \frac {1}{\epsilon^3}  V_1(\frac{|y|}{\epsilon}), \; \;  V_2 ^\epsilon (|y|)=  \frac {1}{\epsilon^2}  V_2 (\frac{ |y|}{\epsilon} )
\end{equation} 

(we omit the index $ m,n $) .

The potential  $ V_3 $ is unscaled. 

Define 

\begin{equation}
 U^\epsilon (|y|) = V^\epsilon_2 +  V_3 \
 \end{equation}
 
  If $ \epsilon > 0$  the Born series converges and the resolvent can be cast in the Konno-Kuroda  form,  [K,K] where the operator $B$ is given as (convergent) power series of convolutions of the potential $ U^\epsilon $ and $V^\epsilon_1 $  with the resolvent of $H_0$.

In general 

\begin{equation} 
\sqrt {V_1^\epsilon  (|y|) + U ^\epsilon (|y|)} \not= \sqrt {V_1^\epsilon  (|y|) }+ \sqrt {U^\epsilon (|y|)  } 
\end{equation} 

and in the Konno-Kuroda  formula for   the resolvent of the operator $H_\epsilon$ one loses separation between the two potentials $V^\epsilon _1 $ and $ U^\epsilon  $.

Notice however that, if $V_1^\epsilon $ and $U^\epsilon$ are of class  $ C^0$ ,   the $L^1 $ norm of $U ^{\epsilon}$ vanishes as $ \epsilon \to 0$ uniformly on the support of $ V^\epsilon _1$. 

By the Cauchy inequality one has 

\begin{equation}
lim_{\epsilon \to 0 }    \| \sqrt { V_1^\epsilon  (y)} .\sqrt { U^\epsilon  (y) }\|^1    = 0 
\end{equation} 

Therefore if the limit exists  the interaction act independently.

We know for [A] that the weak-contact part has a limit in the norm convergence sense.

It follows that the limit of the contact part can be taken in the strong convergence sense.

From [D] one derives that if the limit hamiltonian  represents a unique self-adjoint operator.

This proves theorem 1

\hfill $\heartsuit $

\bigskip

\section{Weak-contact case: separation of the singular part}

Consider now the weak-contact interaction between a particle with a pair of indentical particles.

We allow for the presence of a "regular part" represented by a smooth two body $L^1$ potential of finite range  and call \emph{singular part} the quasi contact interaction and the resonance.

\bigskip

We prove  Theorem 2, that we reproduce here for  convenience of the reader.

\bigskip 

\emph{Theorem 2}

For a  weak-contact interaction of a particle with two identical particles the singular term (pure weak-contact ) and the regular term in the two-body part of the interaction \emph{contribute separately} to the spectral structure of the hamiltonian.

\hfill $ \diamondsuit  $ 

\bigskip

For the proof we uses again  the Konno-Kuroda resolvent formula but now for a system with potentials $ V_2 ^\epsilon + V_3 $. 

Recall that  set  $V_2^\epsilon (|x|) = \frac {1}{ \epsilon^2} V_2 ( \frac {|x|}{ \epsilon })$.

The Konno-Kuroda formula is now for $Re(z )>0 $ and $ R_0^\epsilon (z) = H_0 + \epsilon z $ 

\begin{equation}
 \frac {1}{ H_\epsilon +z} - \frac {1}{ H_0+z} = - \frac { 1}{ H_0+ z } Q^\epsilon B^\epsilon  Q^\epsilon  \frac{ 1}{ H_0+z}  
 \end{equation} 
 
 \begin{equation}
 B^\epsilon  = \sqrt { V_2 ^\epsilon +V_3  } \quad Q^\epsilon ( z)  =  B^\epsilon  \frac {1}R_0(z)  B^\epsilon   \quad R_0 (z) = (H_0 + \epsilon z)^{ -1} 
\end{equation} 

One can now repeat the procedure in Theorem 1.

By assumption $V_2$ and $ V_3 $ are $L^2$ function with finite $ L^2 $ norms and as $ \epsilon \to 0$ 
on the support of $V^\epsilon _2$ the $L^2$ norm of $ V^3  $ is of order $ \epsilon $.

Therefore 

\begin{equation}
\| (\sqrt {V_3 ^\epsilon} + \sqrt {V_2 ^\epsilon })^2 -  V_3 ^\epsilon - V_2 ^\epsilon \| = 0 (\epsilon) 
\end{equation} 

We conclude that in  limit the potentials $V_2 $ and $ V_3 $ contribute \emph{additively} to spectral properties. 

 The potential $V_2 $ (weak-contact) may contribute for a finite or infinite number of elements of the spectrum (depending on the masses and the coupling constants), the potential $V_3 $  gives a contribution to the spectral measure.

In both case there are no singularities at the bottom of the (absolutely) continuous spectrum. 

This proves Theorem 2.

\hfill $\heartsuit $

\bigskip

.........................................

We analyze now the singular part (quasi-contact and zero  energy resonance). 

\bigskip

\emph{Theorem 3 } 

The hamiltonian of a the three-body system of two  identical particles in weak-contact interaction with a third particle is a self-adjoint operator without zero energy resonances.

\bigskip

\hfill $\diamondsuit$

\bigskip

\emph{Proof}

Krein's formula for the resolvent  is 

\begin{equation}
R(z) - R_0 (z) = R_0(z) B^\epsilon  ( 1 - Q ^\epsilon(z) )^{-1})  [ B^\epsilon  R_0 (z)]   \qquad z > 0
\end{equation}  

with

\begin{equation}
R_0 (z) = \frac {1}{ H_0 +z} \qquad Q^\epsilon(z) = B^\epsilon \frac {1}{ H_0 + z} B ^\epsilon \qquad  B^\epsilon= \sqrt{ V^\epsilon  (x)+ V^\epsilon (y)}
\end {equation}

Each $V^\epsilon$ is a sum of two parts:  a singular part that in the limit becomes  concentrated in a point (but "less than a delta function" ) and a \emph{regular part} that has a zero energy resonance. 

We have seen (theorem 2) the in the limit there is no interference between these two types of potentials.

The two weak-contact singularities at $|x_1 - x_3| = 0$ and at $ |x_2 - x_3| = 0 $  originate at $ |x_1 -x_2|= 0 $ a singularity that  for dimensional reasons must be a simple pole. 

If the interaction is sufficiently strong the resulting operator may have a negative spectrum.

The Krein resolvent formulas for the operator converges when $\epsilon \to 0$ and this defines in the limit a self-adjoint operator. 

We will give a detailed analysis in the Appendix but it useful to understand the reason why in the Krein resolvent formula \emph{two resonances do not give a singularity at zero momentum}   (while the same formula in case of a single resonance gives a pole [A]).

When there are two zero energy resonances,  one can write the Birman-Schwinger kernel in two-by-two matrix form.

At zero, the diagonal elements of this matrix are indeed zero, but the off-diagonal elements are not,  since there is no three-body resonance (the potential is a two-body potential).

Therefore \emph{the matrix is not singular at zero} and the Birman-Schwinger series converges uniformly. 

For completeness we give a detailed analysis in an Appendix.

This proves theorem 3.

\hfill $\heartsuit $

\bigskip

Notice that the analysis we have done for the contact or weak-contact interaction can be repeated  subsituting $H_0$ with $H_0+ V $ where $V$ is a regular potential which decays sufficiently fast at infinity,  but the resulting expressions cannot be explicitely given because one does not in general have an explicit expression fo the resolvent.

We have given the details for the case $N=3 $ but the result olds for any $N \geq 3$. 

For a four particle  system the binding of two pairs (to obtain a \emph{ a quadrimer}   is due  to a "residual Coulomb potential" (this is described in [D]). 

In case of weak-contact interaction these "four-body bound states" do not occur.

\bigskip
\emph{Remark}

In the hamiltonian that describes weak-contact interaction of two particles there is no zero energy resonance.

As remarked earlier, the domain of the hamiltonian contains the symmetrized product of the two zero energy resonances. On this vector the Hamiltonian takes value zero. 
 
 If there are no other resonances, as is the case for "pure"  weak-contact interaction, if the parameters (masses and coupling constants) are such that there is a three-body bound state, the Wave Operator for the scattering of a fourth particle off this bound state is bounded in $L^p$ for all $ 1 < p < \infty$.

In particular if the masses $m_i$ are very large the wave function of the bound state has small support
and in the limit the bound state reduces to a fixed point with internal structure. 

\section{The two-dimensional case}  

Also in two dimensions zero range interactions may be of contact or weak-contact type according  to two different scaling of the potental: contact corresponds to the limit as $ \epsilon \to 0$ of the scaling $ V^\epsilon (|x|) =\frac {1}{ \epsilon^2} V ( \frac   { |x|}{  \epsilon})$ while weak contact corresponds to the scaling $ V^\epsilon (|x| =\frac {1}{ \epsilon} V ( \frac  { |x|}{  \epsilon})$.

In the weak contact case, the "potential" is the primitive, with respect to the radial variable, of the two-dmensional delta function.

In order to analyze it, it is convenient, as in the three dimensional case, to introduce the Krein map and find the structure of the potentials in Minlos space. 

One is then led to consider symmetric operator that for three particles is, in the center of mass system ,in the case of weak contact is $ \sqrt {H_0} -C | log (|x_1 --x_2|) $ and in the contact case is $ \sqrt { H_0 } -
\frac {C } { |x-1 -x_2|} $ where $ C > 0$

In the weak contact case the operator in Minlos space is a self-adjoint  operator with possibly a finite number of bound states. 

Inversion of the Krein map gives a self-adjoint operator.with the same number of bound states.

It is a regular perturbation of the two dimensional laplacian. 

In Minlos space, and also in physical space it is the limit, in strong resolvent sense, of operators $ H_0 + V^\epsilon (|x_1 -x_2 |) $ where $ V^\epsilon (|y|) = \frac {1}{\epsilon^2} V ( \frac { |y|}{ \epsilon}$.

In the contact case the image of the operator in Minlos space is the symmetric operator $ \sqrt { - \Delta} - \frac {C}{|y|} \;\; , y \in R^2  \; C > 0 $

There is a constant $C_0 $ such that if $C < C_0$ the operator is positive and extends to a positive self-adjoint operator. 

Inverting the Krein map one has in physical space a positive self-adjoint operator.  

If $C \geq  C_0$ we are in the Weyl limit circe case and in Milnos space there a one-paramter family of self-adjoint; there is a second constant $C_1$ such that if $ C_1 < C \geq C_0$ the operators in the family have a bound state, if $ C \geq C_1 $ each  operator in the family has an infinite number of bound states     with eigenvalues that diverge geometrically (Thomas effect).

Inverting the Krein map one has a family  of self-adjoint operators each of which has an Efimov  sequence of  bound states (the eigenvalues converge geometrically to zero).
with one bound state for extensions 

As in the three dimensional case, to study the convergence of the approximating  hamiltonians $ H^\epsilon = H_0 + V^ \epsilon (|y|) $  (in reference frame of the barycenter) one has to relay on Gamma-convergence.

After having decomposed the self-adjoint operator using the $O(2) $ symmetry (the approximating potentials and the contact hamiltonian are invariant under rotations in the plane) one remarks that the quadratic form of the approximating potentials and of each of the operator in the family are strictly 
convex and that the quadratic forms of the family are well ordered. 

Moreover they are uniformly bounded in $ { \cal H}^{-1} $.

Therefore $ Gamma$-convergence applies [Dal] and there is in the family a unique quadratic form (the lowest laying) that is  limit of the approximating  forms associated to to the approximated  hamiltonians. 

For details and definition of Gamma-converges see [Dal] or  [D] or the brief resume' of [D] given here in a previous section.

Gamma-convergence implies strong resolvent convergence. 

We have proved

\bigskip

\emph{Theorem}

In two dimensions the contact interaction of two particles is a hamiltonian system; the hamiltonian is  a self-adjoint operator which has an Efimov sequence of bound states. 

It is the limit, in strong resolvent sense, of approximating two body  hamiltonians $ H^\epsilon = H_0 + V^\epsilon ( |x_1 -x_2 |) $ where $ V^\epsilon (|y|) = \frac { 1}{ \epsilon^2 } V (\frac {|y|}{\epsilon}) $.

\hfill $ \diamondsuit$

The solutions of the Hamilton equations may have a singular part; the coefficient is called "charge"  and satifies an equation of Volterra type[ C,C,F][ [D,F,T]

\section{Simultaneous  weak contact of three particles}

In two dimensions there is a novel mechanism of production of three-body bound states; the wave function of these bound states has very small support if the masses of the three bodies are very large. 

Consider e.g a system  $S$ made of three particles of mass $m$ in which two are in contact interaction  and a \emph{simultaneous}  weak contact interaction with the third particle (the contact potential is the product of the two). 

\bigskip

\emph{Theorem}

The system $S$  is a hamiltonian system with no zero energy resonances and  an Efimov sequence of bound states. .

\hfill $ \diamondsuit$

\bigskip
\emph{Proof}

Denote by $x_0 \in R^2 $ the coordinates of the third particle and by $x_1, x_:2$ the coordinate of the first two.  

The interaction $  1 \leftrightarrow 2 $ is described by the limit of a two body potential $ V_\epsilon (|x_1 - x_2) = \frac {1}{\epsilon^2 } V( \frac {|x_1-x_2|}{ \epsilon^2 }$ .

The other interaction is described by  the product of two-body central potentials $ V^\epsilon   (x_0 -x_k)  \; k=1,2 $ that scale as $V^\epsilon _k (|x|) = \frac { V _k (|x|)}{ \epsilon }$.

There are no zero energy resonances since  there are identical two-body resonances and the domain of this  hamiltonian can be extended to the symmetric  product of the resonances, an element of $L^2 (R^3 \otimes R^3)$ on which the hamiltonian has value zero. 

For $ \epsilon > 0  $ the \emph{product of the two potentials}  is finite and is zero outside a two dimensional region of area of order $ \epsilon^2 $  and is  rescaled by a factor is of order $ \epsilon^{-2}$ over that region. 

Therefore it converges to a $-C \delta (x) , x \in R^2 $ where $C$ is a positive constant.  

In the limit the hamiltonian represents \emph{a contact interaction} between  particle $1$ and $2$  and a contact interaction particle between particle $0$ and the barycenter of particle $1$ and $2$. 

In  the limit  the interaction tales place  in a point. 

The limit hamiltonian  has no zero energy resonances.  

If this quadratic form is positive it defines a self-adjoint operator; if not it represents an interaction that  has an Efimov sequence of bound states.  .

If the Krein map is inverted  one has a self-adjioint hamiltonian, possibly Efimov bound states. 

\hfill $\heartsuit$

\bigskip

We now describe more in detail the  hamiltonian of the resulting  system. 

To study the structure of the operator  we study its  quadratic form 
 and assume a before that the particles are \emph{ identical bosons} .
 
The wave function  in the frame of reference of the barycenter is best written as  a function of one radial coordinate $r$ and two Euler coordinates on $S^3$.

We define $ r^2 =  (|x_1 - x_3|)^2 +(|x_2 -x_3|)^2 \; \; x_k \in R^2 , \;\; r \in R^+ $.

In the Theoretical Physics literature  this coordinates are called "homogeneous".

The quadratic forms we consider have the same structure as in the case of three dimensions but in two dimension the singularities are different. 

We consider  the case in which a particle is in \emph{simultaneous} quasi-contact interaction with two identical particles. 

This case was discarded in three dimensions because it leads to divergences [M,F].

Since the quadratic form is continuous in  $ \lambda $ at $ \lambda = 0$ one can set $ \lambda= 0$ in equation (3) ; this simplifies the expressions. 

If we denote by $ x_k\in R^2  \;\; k=1,2,3$ the coordinates of the three points with $ x_1 +x_2 + x_0=0$ one has in $ {\cal M}$ for the quadratic forms 

\begin{equation}
Q (\phi) = Q_0 (\phi) + Q_1 (\phi) 
\end{equation}

 where 

\begin{equation}
 \quad Q_0 = (\phi , (H_0+ \lambda) \phi)
\end{equation}

In the center of mass, using Fourier coordinates conjugated with $x_1- x_3$ and $ x_2 -x_3$ ,  the \emph{kernel} of $ Q_1$  is 

\begin{equation} 
\frac{1} { (q_1^2 + q_2^2 + (q_1,q_2  )) ( q_1 +q_2) ^2 } \qquad q_i \in R^2
\end {equation} 

Setting  

\begin{equation}
( x_1-x_3)^2 + (x_2-x_3) ^2 = r^2 \quad r \in R^+ 
 \end{equation}

 the kernel $Q_1$  can be written in spacial  homogeneous coordinates  as integral over $S^3 $ of a kernel that in the radial coordinate has a singularity $ -C \frac { 1} { r }$.

On now proceeds as in the three body case in $R^3 $ with contact interactions.

But notice that now the system is made of three particles in two dimensions. 

If the particles are identical in the center-of mass frame the hamiltonian has the form $ H_0 - \frac {c}{r} $ 
where $H_0 $ is the free hamiltonian for three-particle hamiltonian with mass $m$ in $R^2 $ and $r$ is defined in (23).

This hamiltonian  is a self-adjoint operator in $ L^2 (R^2 \otimes R^2 ) $ which  has a number of bound states that increase when the masses  of the three bodies became very large (the free hamiltonian \emph{decreases}. 

It describes the system in the center-of-mass reference frame. 

For each finite value of the mass $m$  it is the limit, \emph{in strong resolvent sense}, of   hamiltonians 

\begin{equation}
 H_0 + \sum_{k=1}^2 V_k \qquad    V_1^\epsilon (|y|) = \frac {C_1 }{ \epsilon^2  } V( \frac {|x_2-x_1|}{\epsilon}) \qquad  V_2^\epsilon = \frac{ C_2 }{ \epsilon^2} V(\frac  { |x_2 -x_0|} {\epsilon}) V (\frac  {| x_1 -x_0|} {\epsilon}) 
 \end{equation} 

\bigskip

\emph{Remark} 

If one takes $m = \frac {1}{\epsilon} $ and one chooses $\epsilon $ to be the same parameter that us used to approximate the contact interaction, the  limit hamiltonian  for $ \epsilon \to 0  $ describes an interaction   that takes place in a point.

Therefore in the limit  the interaction with a fourth particle is a "point interaction" (it is concentrated in one point) and  since there are no zero energy resonances has a Wave Operator that is a bounded map from $L^p$ to  $L^p$ for all $ 1 < p < \infty$.

\bigskip
............................

\bigskip

A system with the same properties has been described in [C,M,Y]

\section{Acknowledgements} 

I benefited much form correspondence with the late R.Minlos. 
 I'm very grateful to A.Michelangeli for relevant comments at an early stage of this research. The appendix of the present paper is joint work with him (unpublished).

\bigskip

\section{References}

[A]  S.Albeverio, F.Geztesy, R.Hoegh-Krohn, H.Holden Solvable models in Q.M. AMS 2004

[A,S]   Alonso, B.Simon J. Operator Theory  4 (1980) 251-270 ;  6  (1981)  407

[C,M,Y] H.Cornean , A.Michelangeli K.Yajima   arXiv 1804.01297

[C,C,F]  R.Carlone, M. Correggi, R.Figati  Funcl Anal.  Op. Th. for Q. Physics (2017) 189-211

[D] G.F. Dell'Antonio   arXiv 1804.06747

[D,F,T] G.Dell'Antonio, R.Figari, A.Teta   Ann. I. H. Poincar 60 (1994) 253.290

[D,R]   S.Derezinski, S.Richard ArXiv: 1604.03340

[Dal] G.Dal Maso Introd. to Gamma convergence Progr.Nonlin.Diff.Eq. 8, Birkhauser (1993)

E] V. Efimov Phys. Letters 33B (1970) (1970) 563-566

[E,G,G] B.Erdogan, M.Goldberg , W.Green   arXiv 1706.01530

[K,K] R.Konno, S.T.Kuroda  Journ, Jap. Math, Soc. 1966 55-63

[KS] W. Klaus, B.Simon  Ann.Inst. Poincar\'e  XXX (1979) 83-87

[M1] R. Minlos Int.Scol. Reseach network  2012  article ID 230245 . 

[M2] R.Minlos Moscow Math.Journal  14 (2014) 617-637

[M,F] R.Milnos, L.Faddaev  Sov. Phys. Doklady 6 (1972) 1072-1074

[ST] G.Skorniakov, K. Ter-Martirosian Sov. Phys. Jetp 4 (1957) 648- 654

\bigskip 
\section{Appendix: Details for the pure quasi-contact case.}

For a rigorous analysis one can proceed as in the two-particle case [A] using the same unitary operators $ U^\epsilon $ as in [A] to change (artificially) the scale.

W denote the potential by $ - V((|x|) \;\; V> 0 $.

Notice that this change of scales provides an extra factor $ \epsilon$ in the Krein resolvent formula
 and the volume of the support of $ V^\epsilon $ is now of order $ \epsilon^3 $ (as in the contact case). 
 
 Up to an error of the order of $ \epsilon^3 $ we have for $z >0 $

 $$
 \frac {1}{ H^\epsilon +z } - \frac {1}{ H_0+z} = - \frac {1}{ H_0+z} [ \sqrt {V^\epsilon(x)} K_\epsilon  \sqrt {V^\epsilon(x)  }  +
 $$
 \begin{equation}
  \sqrt V"\epsilon (x) K^\epsilon \sqrt {V^\epsilon(y)}  + \sqrt {V^\epsilon(y)} K^\epsilon \sqrt {V^\epsilon)} + \sqrt {V^\epsilon(y)}  K^\epsilon \sqrt {V^\epsilon(y)} ] \frac {1}{ H_0+z}  
\end{equation}
where 
\begin{equation}
K^\epsilon = \frac {1} { 1 - \sqrt {V^\epsilon (x)  + V^\epsilon (y)}  \frac {1}{ H_0+z } \sqrt {V^\epsilon (x)  + V^\epsilon (y)}  }  
\end{equation}
 
There are  four terms two \it diagonal \rm and two \it off-diagonal.\rm 

We begin analyzing the first diagonal term. 

To control the scaling we introduce as in [A] two partial scaling operators in $ L^2 (R^3 \times R^3)  $

\begin{equation}
U^\epsilon F(x,y) = \frac {1} { \epsilon^{ 3/2} }F( \frac { x}{ \epsilon}, y ) 
\qquad 
V^\epsilon F(x,y) =  \frac {1} { \epsilon^{ 3/2} } F(x,  \frac { y}{ \epsilon} ) 
\end{equation}

Under this scaling one has

\begin{equation}
(U^\epsilon)^* H_0 U^\epsilon ={1 \over \epsilon^2 }
 - { m+1 \over  2m} \Delta_x - \epsilon^2 {m+1 \over 2m} \Delta_y - \epsilon {1\over  m} \nabla _x \nabla_y    \end{equation}

where $m$ is the reduced mass of ht esystem. 

Therefore

\begin{equation}
(U^\epsilon)^*  ( H_0 +z) U_\epsilon = { 1 \over \epsilon^2}   { m+1\over 2m} \Delta_x - \epsilon^2 {m+1 \over  2m} \Delta_y - \epsilon {1 \over m} \nabla _x \nabla_y + \epsilon^2  z    
\end{equation}

Inserting the identity  $ 1= U^\epsilon (U^\epsilon)^*$ and redistributing power of $ \epsilon $ one has

\begin{equation}
\frac {1} {H_0+z} \sqrt { V^\epsilon } K^\epsilon \sqrt { V^\epsilon } \frac {1} { H_0+z} \equiv  A_\epsilon B_\epsilon C_\epsilon 
\end{equation}

where  

\begin{equation} 
A^\epsilon (x,x'; y,y') = G_z (x-\epsilon x') \sqrt { V _\epsilon (x') } ,\qquad C^\epsilon = \sqrt { V(x')}  G_k (\epsilon x - x'); y-y')  
\end{equation}

where $ G_z $ is the kernel of $ ( H_0 +z )^{ -1} $.

From this we deduce for $ \epsilon \to 0 $ strongly 

\begin{equation}
A^\epsilon \to A, \qquad C^\epsilon \to C  
\end{equation}

where 

\begin{equation}
A(x,x'; y,y') = G_z (x, y-y')   \sqrt {V(x')}   \qquad C(x,x'; y,y') = \sqrt { V(x') } G_z (-x'; y-y')  
\end{equation}
 
We now take into account that the hamiltonian $ H_0 - V^\epsilon (x)$ leaves invariant the space of functions that do not depend on the variable $y$ and acts there as the hamiltonian of the two-body problem. 

In particular it has a zero energy resonance i.e. a function of two vector-valued coordinates  which has the form $ \tilde \psi (x,y) = \psi (x) $  where $ \psi (x) $ is the resonance of the two-body problem. 

This simplifies the problem because we can use the analysis made for the two-body problem. 

In the diagonal terms one scales only with respect to one of the variables.

This cancels (to first order in $ \epsilon )$  the terms in the denominator that contain gradients with respect to the other variable and the part of the potential that depends on the other variable. 

 The remaining term is the hamiltonian in one of the two channels plus the laplacian in the other channel;  
 
 This operator admits as resonance function , the resonant function in the given channel times the identity in the other variable.
 
The two-body resonances \emph{contribute separately} in the limit formula for the resolvent; this is due to the fact that the two-body resonances are of the type $ \phi (x_1 - x_3 )\otimes  1 $ and $1\otimes \phi  (x_2 -x_3) $ where $ \phi $ is the two-body resonance function.

Define 
\begin{equation}
W(z) =lim_{\epsilon \to 0 }W^\epsilon (z)  \qquad W^\epsilon (z) =  \frac {1 }{ H^\epsilon + z }- \frac {1}{ H_0 + z}    \quad Re z > 0 
\end{equation}

Then  
\begin{equation} 
W (z) =  A ( \frac { | \psi > < \psi | }{ \frac { i \sqrt z}{ 4 \pi} |<\sqrt V . \psi> |^2 } \otimes I_y ) C + 
A' ( I_x \otimes \frac { | \psi > < \psi| }{ \frac { i\sqrt z  }{ 4 \pi} |<\sqrt V . \psi> |^2 } ) C '  
\end{equation}
where 
\begin{equation}
A(x,x'; y,y') = G_z (x;y-y') \sqrt { V(x')} ,\qquad C(x,x';y,y') = \sqrt{ V(x) } G_z(-x'; y-y')  
\end{equation} 

and analogous formula for $A', \; C' .$

As before wee have denoted by $ G_z (x,y) $ the kernel of $ \frac {1} { H_0 +z } $ 

As in the two-body case we normalize the resonance by $ <V, \psi>  = 1 $ and so the projecton is on a unit vector which depends on the potential  $V.$

This shows that the limit  is a positive  symmetric operator.

Therefore it admits a unique self-adjoint extension. 

The domain of the extension consists of the completion in $L^ 2(R^6)$ of the Sobolev space $ { \cal H}^1 \otimes {\cal H^1} $ with  linear combination of the function of the form $\phi (x) \xi (y) + \xi (x) \phi (y) $ where $ \phi \in L^2 (R^3)$ and $ \xi $ is the zero energy resonance. 

Denote by $ S_z $ the limit $ \epsilon \to 0$. 

In order to prove that this limit provides the resolvent of the  self-adjoint operator $ H$ one must still prove the identity  

\begin{equation}
 (\frac {1}{ H_0 +z }  + W (z) ) (H+z)=  (H+z)(\frac {1}{ H_0 +z}  + W_ z )= I    
\end{equation} 

where $W(z)$ is defined in eq. 37 and $ H = \lim_{\epsilon \to 0} H^\epsilon $.

Since in this case one has the explicit expressions of the kernels and the power series converges strongly  one can verify with a straightforward analysis that this relation is indeed satisfied as a relation between quadratic forms on a dense domain. 

\end{document}